\documentclass[aps,twocolumn,prl,color,psfig,epsf,sort&compress,floatfix]{revtex4-1}
\usepackage{amssymb}
\usepackage{amsmath}
\usepackage{bm}
\usepackage{soul}
\usepackage[normalem]{ulem}
\usepackage[dvipsnames]{xcolor}

\usepackage{hyperref}
\usepackage{graphicx}
\usepackage{etoolbox}
\usepackage{tikz}

\newrobustcmd*{\mysquare}[1]{\tikz{\filldraw[draw=#1,fill=#1] (0,0)
rectangle (0.2cm,0.2cm);}}

\begin{document}

\title{Yield-stress transition in suspensions of deformable droplets}

\author{Giuseppe Negro$^{1}$, Livio Nicola Carenza$^{2}$, Giuseppe Gonnella$^{1}$, Fraser Mackay$^{3}$, Alexander Morozov$^{3}$, Davide Marenduzzo$^{3}$}

\affiliation{
$^1$Dipartimento di Fisica, Universit\'a degli Studi di Bari and INFN, Sezione di Bari, via Amendola 173, Bari, I-70126, Italy, \\
$^2$ Instituut-Lorentz, Universiteit Leiden, P.O. Box 9506, 2300 RA Leiden, Netherlands,\\
$^3$ SUPA, School of Physics and Astronomy, University of Edinburgh, Peter Guthrie Tait Road, Edinburgh, EH9 3FD, UK 
}

\begin{abstract}
Yield-stress materials, which require a sufficiently large forcing to flow, are currently ill-understood theoretically. To gain insight into their yielding transition, here we study numerically the rheology of a suspension of deformable droplets under pressure-driven flow. We show that the suspension displays yield-stress behaviour, with the droplets remaining motionless when the applied body-force is below a critical value. In the non-flowing phase, droplets jam to form an amorphous structure, whereas they order in the flowing phase. Yielding is linked to a percolation transition in the contacts of droplet-droplet overlaps, and requires suitable wetting boundary conditions and strict conservation of the droplet area to exist. %Relaxation of the latter assumption leads to droplet flow at any forcing due to evaporation-condensation. 
Close to the yielding transition, we find strong oscillations in the droplet motion which closely resemble those found experimentally in confined colloidal glasses under flow. %Finally, 
We show that even when droplets are static the underlying solvent moves by permeation, so that the viscosity of the composite system is never truly infinite, and, as we discuss, its precise value ceases to be a bulk material property of the system.
\end{abstract}

\maketitle

Yield-stress fluids are materials which flow only when subject to a sufficiently large stress, or external forcing~\cite{Jones2002,Bonn2017}. The critical stress above which there is flow is known as the yield stress. Examples of yield-stress fluids abound in everyday materials and include toothpaste, whipping or shaving cream, mayonnaise and cement. The defining property of an ideal yield-stress fluid is that the apparent viscosity should be infinite below yielding, so that the yield stress should mark a transition between a solid-like and a fluid-like regime. Nevertheless, in practice it is often arduous to distinguish this behaviour from that of a strongly shear-thinning fluid for which the viscosity drops by orders of magnitude at the yielding point, such that the material always flows albeit very slowly under any external forcing, however small~\cite{Bonn2017}. 
%maybe split sentence at the end

Phenomenological theories for yield-stress fluids typically assume a non-Newtonian and non-linear relation between the shear stress $\sigma$ and the shear rate (or velocity gradient) $\dot{\gamma}$. A popular model is the Herschel-Bulkley one~\cite{Herschel1926}, which is based on the generic equation $\sigma=\sigma_y+\eta_{\rm \infty}{\dot{\gamma}}^n$, with $\sigma_y$ the yield stress, $\eta_{\infty}$ a material parameter and $n$ a generic exponent found by fitting experimental data, and smaller than $1$ for shear-thinning fluids. Phenomenological models like this are extremely useful to analyse and compare experiments, but -- by their nature -- they do not address the fundamental physical mechanisms underlying the {\it existence} of a yield stress. 

%best described by contrasting it with a Newtonian fluid such as water or honey. In a Newtonian fluid, the shear stress $\sigma$ is linearly proportional to the shear rate (or velocity gradient) $\dot{\gamma}$: $\sigma=\eta \dot{\gamma}$, with $\eta$ the viscosity of the material, which is a constant. In a yield-stress fluid, first $\dot{\gamma}=0$ if $\sigma\le \sigma_y$, with $\sigma_y$ the yield stress. Then, if $\sigma>\sigma_y$, the fluid flows and its rheology is phenomenologically described by $\sigma=\sigma_y+\eta_p \dot{\gamma}^n$, where $\eta_p$ is the effective viscosity at infinite forcing ($\dot{\gamma}\to\infty$) and $n$ is an exponent: the Bingham model assumes $n=1$, whereas in the more general Herschel-Bulkley model $n$ is unspecified, and is typically between $0$ and $1$.

Yield-stress fluids can be characterised according the softness of their constituents~\cite{Bonn2017,Zac2}, and range from bubble foams~\cite{Heller1987,Zinchenko2017,Cloitre2003} to suspensions of nearly-hard colloidal spheres~\cite{Pusey1986,Pham2006} (e.g., spherical particles stabilised sterically with a thin polymer layer). In all cases, at large enough particle concentrations -- such that the system is the jammed, or glassy, phase respectively -- these materials are experimentally known to undergo a yielding transition. They also display soft glassy rheology, as described by the Herschel-Bulkley model~\cite{Sollich1997,Holmes2003,Paredes2013}. In colloidal fluids, rheological experiments further show that the effective viscosity of the system becomes very large, and possibly diverges~\cite{Bonn2017,Pham2006}. A confounding factor hampering a conclusive demonstration of ideal yield-stress behaviour in experiments is that the solid-like phase in a colloidal glass is amorphous, and the fundamental physics of the amorphous state is not fully understood~\cite{Bonn2017}. Additionally, as we show to be relevant here, colloidal glasses or foams are composite materials, so that the behaviour of the dispersed particles and the underlying solvent may differ, thereby complicating the picture.

Here we consider a generic universal model system for a yield-stress fluid: a suspension of soft deformable droplets embedded in a Newtonian fluid~\cite{Foglino2017,Foglino2018,Tiribocchi2020,Zac1}. By changing the particle softness -- in particular interpenetrability and surface tension -- our model encompasses foams, stabilised oil-in-water emulsions and colloidal suspensions as special cases. We show that these deformable suspensions display the hallmark of yield-stress fluids, as the droplets are immobile even when subjected to a (small) pressure difference, or forcing. In this immobile phase, the droplets are arranged in an amorphous pattern, and the network of droplet-droplet contacts, or overlaps, percolates.
These overlaps provide the soft analogue of frictional contacts, which are known to play a crucial role in colloidal rheology~\cite{Wyart2014}. Upon yielding, contact percolation is lost, while droplets order as they flow. Importantly, we find that even in the phase where droplets are static the solvent flows by permeation, meaning that the viscosity of the overall system is never truly infinite. Close to the yielding point, sustained velocity oscillations occur, similarly to what found experimentally in flowing colloidal fluids close to the glass transition~\cite{Isa2009}. Our results %deformable suspensions provide a paradigmatic model for the hydrodynamics of yielding in composite colloidal materials, which 
allow to gain more insight into the microscopic physical mechanisms underlying the yielding transition. In our system, the latter is controlled by an inverse Bingham number, measuring the ratio between viscous and interfacial forces. Notably, this is the same number which controls discontinuous shear thinning at larger forces~\cite{Foglino2017,Foglino2018,Fei2020}. Finally, our simulations suggest that yielding disappears when the droplets area is not strictly conserved, suggesting that systems featuring evaporation-condensation phenomena can evade yield-stress behaviour and flow under any forcing. 
%We argue that such oscillations have a synchronizing effect on the motion of the single units leading to a percolating state in the direction defined by the pressure gradient.
%Notably, the yielding transition disappears when the droplet area can fluctuate, suggesting that systems featuring evaporation-condensation phenomena can evade yield-stress behaviour and flow under any non-zero pressure gradient.
%Our simulations allow us to dissect the different responses of the droplets and underlying solvent, and show that even in the solid-like phase the solvent flows by permeation, meaning that the viscosity of the system is never truly infinite.  When the yielding point is approached, large and periodical velocity oscillations occur, just as found experimentally in flowing colloidal fluids close to the glass transition~\cite{Isa2009}. 
%Our system may provide a generic tractable paradigm for the hydrodynamics of yielding in composite colloidal materials.

%This model is also appealing as lubrication forces, which are notoriously challenging to accurately account for in simulations~\cite{Nguyen2002,Wyart2014} are altogether absent. Nevertheless, frictional forces, which are known to play a crucial role in colloidal rheology~\cite{Wyart2014}, are present, in the form of droplet-droplet overlaps

\begin{figure*}[ht!]
  \includegraphics[width=1.9\columnwidth]{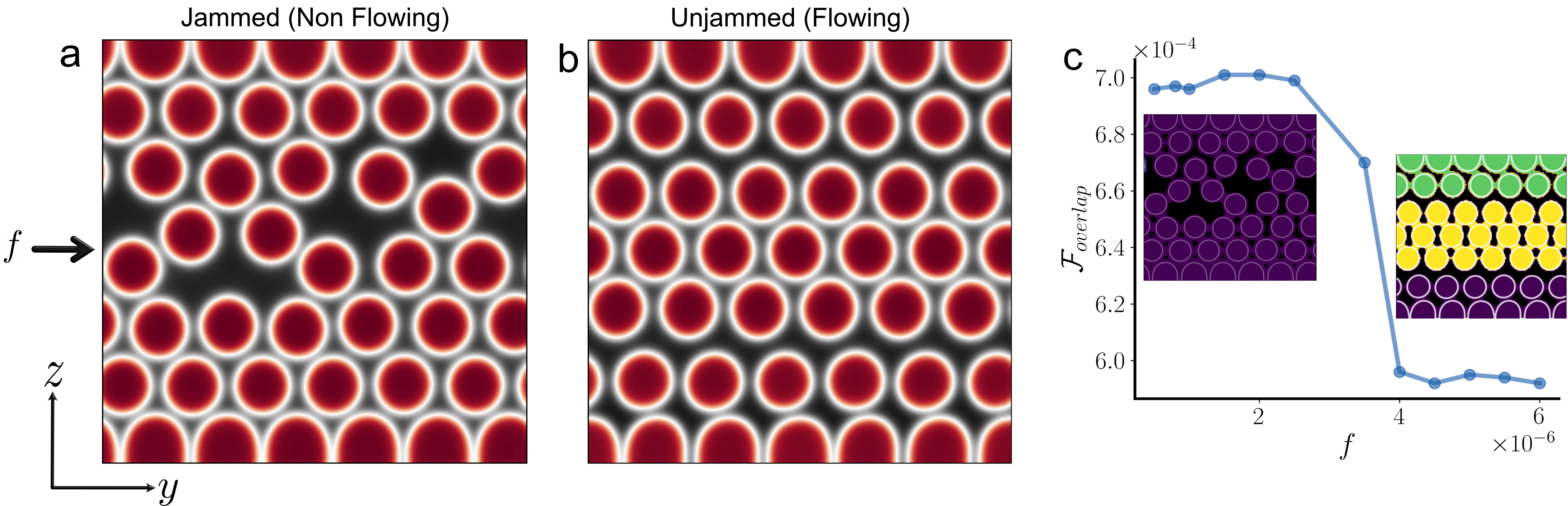}
  \caption{\textbf{Yielding transition in the conserved model.} \textbf{(a)-(b)} Color map of $\phi=\sum_{i}\phi_i$ for $f<f_c$ ($f=2.0\times 10^{-6}$ in panel (a)) and $f>f_c$ ($f=4.0\times 10^{-6}$ in panel (b)), for the conserved model. Black and red regions correspond to $\phi=0$ and $\phi=2$ respectively.  \textbf{(c)} Free energy of overlaps (see text)  as a function of body-force $f$. The insets of panel (c) show clusters of contacting droplets, resulting from a density-based spatial clustering analysis on the free energy of overlaps~\cite{SI}. Different colors correspond to different clusters. Left and right inset correspond to the configuration shown in panels (a) and (b) respectively. Movies of the dynamics corresponding to (a) and (b) can be seen in Suppl. Movies 1 and 2 respectively~\cite{SI}.}
  \label{fig1}
\end{figure*}
%, but only controlled via a Lagrange multiplier. 
%%%%%%%%%%%%%%%%%%%%%%%%%%%%%%%%%%%%%%%%%%%%%%%%%%%%%%%%%%%%%%%%%%%%%%%%%
%\emph{We track down this fundamental difference and show it is due to the different velocity-force curves for a single droplet wetting the wall with conserved or non-conserved dynamics. \textcolor{red}{(What is a nonsingle droplet? And more importantly, said like this it looks like the yielding transition is only due to boundaries... (are we saying that the emulsion of (conserved) droplets would eventually flow if we considered a Kolmogorov rather than Poiseuille flow?))}} \emph{For a system with strict volume conservation, there are also large velocity oscillations close to the yielding transition, just as found experimentally in flowing colloidal fluids close to the glass transition~\cite{Isa2009}}.  
%%%%%%%%%%%%%%%%%%%%%%%%%%%%%%%%%%%%%%%%%%%%%%%%%%%%%%%%%%%%%%%%%%%%%%%%%

%is the yield stress real? is it linked to a solid-to-liquid transition or does it separate two regions with very different viscosity?~\cite{Bonn2017}.
%what can we say on this? the particles are stuck but does the fluid flow? also wetting and evaporation-condensation leads to evading the yield stress in simulations with non-conserved order parameters. 

%methods

To study the rheology of our soft droplet suspension, we work in $2D$ (Fig.~\ref{fig1}) and consider two models: the first strictly conserves the area of each droplet, the second allows it to fluctuate around a target value, for instance due to evaporation or condensation phenomena. We refer to these as the conserved and non-conserved model, respectively. In both cases, the $N$ droplets in the system are non-coalescing, and we ensure this by describing them in terms of $N$ distinct phase fields, $\phi_i$, with $i=1,\ldots, N$. The hydrodynamics of the suspension can then be studied by following the coupled evolution of the phase-field variables and of the velocity field $\mathbf{v}$ of the underlying solvent. The use of phase-field means that lubrication forces, which are notoriously challenging to accurately account for in simulations~\cite{Nguyen2002,Wyart2014} are altogether absent. At the same time, frictional forces, which are known to play a crucial role in colloidal rheology~\cite{Wyart2014}, are present, in the form of overlaps between different phase-fields.

The thermodynamics  of the conserved model is governed by a free energy $\mathcal{F}$  whose density is
\begin{equation}
\label{freeenergy}
\sum_i^{N} \frac{\alpha}{4}\phi_i^2(\phi_i-\phi_0)^2 + \frac{K}{2}\sum_i^{N}(\nabla\phi_i)^2  + \sum_{i,j,i<j}\epsilon\phi_i^2\phi_j^2.
\end{equation}
Here, the first two terms favour the formation of circular droplets with $\phi_i\simeq \phi_0$ in their interior, and $\phi_i\simeq 0$ outside. 
%They 
The material constants $\alpha$ and $K$ determine the surface tension $\gamma=\sqrt{8K\alpha/9}$ and the interfacial thickness  $\xi=\sqrt{2K/\alpha}$ of the droplets~\cite{Pagonabarraga2002}. The term proportional to $\epsilon>0$ describes soft repulsion  pushing droplets apart when overlapping. The phase-field variables evolve according to a set of coupled Cahn-Hilliard equations, 
\begin{equation}
\label{cahn-hilliard}
\frac{\partial \phi_i}{\partial t} + \mathbf{v} \cdot \mathbf{\nabla} \phi_i = M\nabla^2 \mu _i 
\end{equation}
where $M$ is the mobility, $\mu_i=\delta {\cal F}/\delta \phi_i$ the chemical potential of the $i$-th droplet.
%, and ${\cal F}$ the total free energy. 
The flow obeys the Navier-Stokes equation
\begin{equation}
\label{navier-stokes}
\rho \Big(\frac{\partial}{\partial t} + {\mathbf v}\cdot {\nabla}\Big){\mathbf v} = -{\nabla p} + \mathbf{f}^{\rm th} + \eta_0\nabla^2{\mathbf v}+f\mathbf{e_y},
\end{equation}
where $\rho$ indicates the total fluid density, $p$ denotes the hydrodynamic pressure and $\eta_0$ the solvent viscosity~\cite{SI}. The term $\mathbf{f}^{th} = -\sum_i \phi_i{\nabla}\mu_i$ stands for the internal thermodynamic force field due to the presence of non-trivial compositional order parameters, while $f$ is the magnitude of the body-force,  %$\mathbf{f}$ (whose magnitude we call $f$ \Livio{[then why don't we use the boldmath $\bm{f},\bm{f}^{th}$ since the beginning and avoid this remark? (I would also write $\mathbf{v}$ as $\bm{v}$.]}) is the body-force, 
which we take along the horizontal direction (Fig.~\ref{fig1}a).
%and as such it can also be expressed as a divergence of a stress tensor~\cite{Cates2017}. The 

In our second model for the non-conserved concentration field, the free energy ${\cal F}$ is supplemented by an additional term,
\begin{equation}
\label{constraint}
\mathcal{F}_{\rm constraint}= \lambda\left(1-\frac{1}{\pi R^2 \phi_0^2}\int  dydz\,\phi_i^2\right)^2,
\end{equation}
%droplet areas are conserved approximately, rather than strictly. 
%The free energy is given by the spatial integral ) with an additional soft constraint, 
with $\lambda>0$ a constant which quantifies droplet compressibility, and provides a soft constraint for the droplet area.
%deviation of the $i$-th droplet area from the target value, the area of a disc of radius $R$. To write down the phase-field dynamical equations, we assume simple relaxational and overdamped dynamics,
The phase-fields evolve according to a relaxational and overdamped dynamics defined by
\begin{equation}\label{phieq}
\frac{\partial \phi_i}{\partial t} +\mathbf{v}\cdot\nabla \phi_i= - \frac{1}{\Gamma}\frac{\delta \mathcal{F'}}{\delta \phi_i}, 
\end{equation}
where $\Gamma$ is a friction-like parameter and $\mathcal{F'}=\mathcal{F}+\mathcal{F}_{constraint}$. The %Navier-Stokes 
equation for $\mathbf{v}$ 
is still given by Eq.~(\ref{navier-stokes}).

The dynamics are integrated with a parallel hybrid lattice Boltzmann approach~\cite{Tiribocchi2009,Marenduzzo2007,Carenza2019} where Eq.~\eqref{navier-stokes} is solved by a lattice Boltzmann algorithm, and Eqs.~\eqref{cahn-hilliard},\eqref{phieq} are solved by finite difference methods. We consider flow in a channel with no-slip boundary conditions at the top and bottom walls, driven by a fixed pressure difference along the $y$ direction  -- leading to Poiseuille, parabolic, flow for a Newtonian fluid. At the wall, neutral wetting boundary conditions are imposed for each droplet. For more details, and a full list of parameters used, see~\cite{SI}. %Values of $\Delta p$ in what follows are given in simulation units. While the trends we discuss are generic, the simulations we report can be mapped to a system with $\sim 100\mu$m-size droplets whose surface tension is $\gamma\sim$mN/m (see SI), embedded in a background Newtonian fluid with viscosity $\eta_0=10$ cP. Our model differs from that used in~\cite{Falcucci2008} to study the glassy dynamics of foams and sprays, which in general allows for droplet coalescence. 

%results

%Key results

%1. depinning transition single droplet in conserved model but not in non-conserved model, link to evaporation-condensation

%2. yield stress in conserved model but not in non-conserved model

%3. oscillations close to the yielding transition

\begin{figure}[t!]
  \includegraphics[width=\columnwidth]{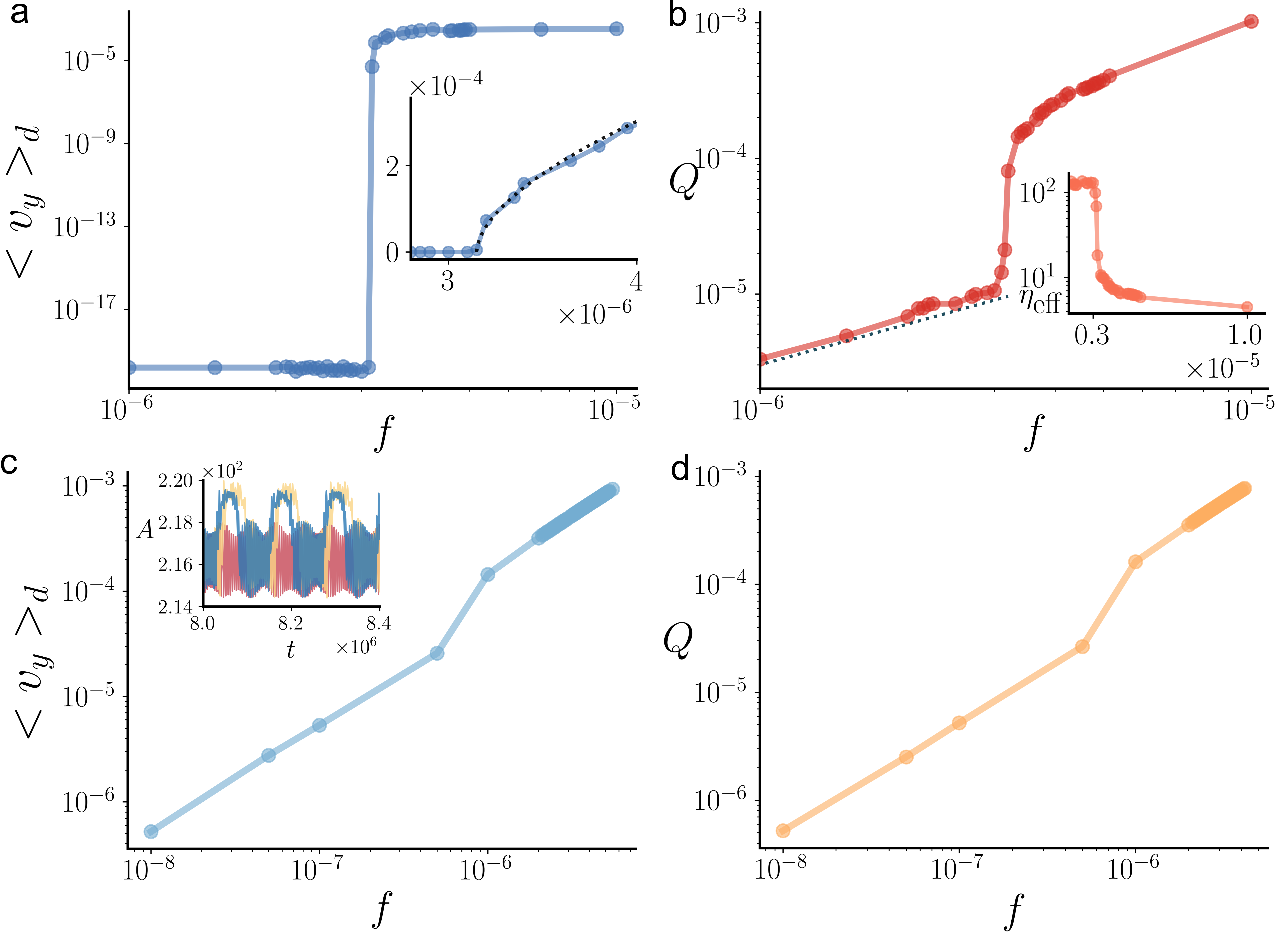}
  \caption{\textbf{Flow behaviour in the conserved and non-conserved models}. \textbf{(a)-(b)} Average droplet velocity (a) and throughput flow (b) for the conserved model. The inset of panel (a) shows the mean droplet speed close to criticality and the result of the fit (dashed line) with the function $\langle v\rangle \propto (f-f_c)^{\beta}$, with $\beta\simeq 0.54$. The inset of panel (b) shows the effective viscosity $\eta_{\textrm{eff}}$ as a function of the body-force $f$.  \textbf{(c)-(d)} Average droplet velocity (c) and throughput flow (d) for the non-conserved model.  The inset of panel (c) shows the area of three nearby droplets versus time for $f=1.0\times 10^{-6}$. }\label{fig2}
\end{figure}

%links to percolation/force chains, link to frictional interactions, here touching due to the overlap between droplets?
%\textcolor{red}{Va stressata la novita del fatto che facciano yielding!!!!! Quale è il risultato più importante. Il modello è paradigmatico perche cià molte cose. Non si capisce che cosa è nuovo rispetto a quello che è gia stato fatto.}
We first study the rheological response of a droplet suspension (with packing fraction $\varphi\simeq 0.5$) in the conserved model. A key result is that there exists a critical body-force $f_{c}$ separating two fundamentally different behaviours. 
For small forcing (Figs.~\ref{fig1}a and \ref{fig2}a) the suspended droplets are jammed and settle into a stationary non-flowing configuration (Suppl. Movie 1) where they are immobile for the whole duration of the numerical experiment ($\sim \mathcal{O}(10^8)$ iterations). The snapshot shown in Fig.~\ref{fig1}a shows a typical  droplet configuration for this regime. For larger $f$, there is a subtle morphological rearrangement of the droplets (Fig.~\ref{fig1}b), which is accompanied by a yielding transition, as droplets now steadily move (Fig.~\ref{fig2}a). The snapshot shown in Fig.~\ref{fig1}b shows a typical late time configuration, which is travelling from left to right at a fixed velocity (Fig.~\ref{fig2}a and Suppl. Movie 2).
%In particular, 
An inspection of the configurations shows that while the non-flowing state is amorphous (Fig.~\ref{fig1}a), in the flowing state droplets order (Fig.~\ref{fig1}b; see~\cite{SI} for a quantification of flow-induced ordering).
%\Livio{In order to quantitatively characterize such morphological transition we recall that the variance of the distribution of the $N$ suspended droplets  $var(N)  \sim L^\alpha $. Uniformly distributed systems have $\alpha=d$, with $d$ the system dimension. Systems with broken positional degrees of freedom as crystals and quasi-crystals have suppressed density fluctuations and $\alpha<d$. The behavior of the non-universal exponent $\alpha$ measured on the configurations across the transition in Fig. xxx shows that the system becomes more and more hyperuniform as $\alpha$ decreases as the applied body-force increases. We can rationalize  
This morphological adjustment is accompanied by a fundamental change in the patterns of contacts, or overlaps, between droplets. As shown in the left inset of  Fig.~\ref{fig1}c, such overlaps create a percolating network in the non-flowing state, whereas after yielding contacts no longer percolate along the flow gradient direction (right inset of Fig.~\ref{fig1}c and Fig S1). The change in droplet contacts can be quantified by plotting the overlap free energy ($\mathcal{F}_{\textrm{overlaps}}=\epsilon\int  dydz \sum_{i,j}\phi_i^2\phi_j^2$) as a function of body-force (Fig.~\ref{fig1}c): this quantity drops sharply at the yielding transition, corresponding to the loss of contacts between droplets near the wall (right inset). As discussed in more detail below, another key feature is that droplets need to deform at least transiently when the system yields~\cite{SI}. 

To quantify the yielding transition we compute two quantities: \emph{(i)} the mean velocity of the droplets' center of mass $ \langle v_y \rangle_d, $ (Figs.~\ref{fig2}a,c) and \emph{(ii)} the throughput flow $Q=\int dydz \  \textrm{v}_y$ (Figs.~\ref{fig2}b,d). 
While  $\langle v_y \rangle_d$ quantifies the motility of the suspended particles, $Q$ can be used to compute the effective viscosity of the suspension, $\eta_{\rm eff}$.  The latter quantity can be estimated as $\eta_{\rm eff}=\eta_0 \frac{Q_0}{Q}$, where $Q_0=\frac{f L^3}{12\eta_0}$ is the throughput flow of a Newtonian fluid with viscosity $\eta_0$ subject to a body-force $f$, and $L$ the channel width. 
The yield-stress behavior is apparent from the plot of $ \langle v_y \rangle_d$ in Fig.~\ref{fig2}a. Close to criticality, the mean droplet speed behaves as $ \sim (f - f_c)^{\beta}$, with $\beta\simeq 0.54$ (dotted curve in the inset of Fig.~\ref{fig2}a). The phenomenology resembles that of the Prandtl-Thomlinson model (where $\beta=1/2$) which describes a particle in a dashboard potential, and provides a simple microscopic model for dry friction~\cite{Popov2017}. %and depinning~\cite{Popov2017}.

Importantly, even in the non-flowing phase in which droplets are at rest, the underlying solvent flows (Fig.~\ref{fig2}b): indeed $Q$ is non-zero for all values of $f$. In more detail we find that there is a well-defined linear regime at small forcing, which corresponds to a high but finite effective viscosity (inset of Fig.~\ref{fig2}b). In stark contrast, yield-stress fluid under shear exhibit wall slip and an infinite effective viscosity~\cite{Bonn2017}. This shows the exact value of $\eta_{\rm eff}$ depends on the geometry of the system, and hence can no longer be viewed one of its bulk material property. %Indeed, $\eta_{\rm eff}$ in our pressure-driven flow depends on system size $L$~\cite{SI}.
%The latter quantity can be defined by dividing the throughput flow of a Newtonian fluid with viscosity $\eta_0$, given by \textcolor{red}{$Q_0=\frac{f L^3}{12\eta}$}, \textcolor{red}{with L the system size}, by the throughput flow of the suspension, $Q$. 
The flow at $0< f< f_c$ is purely permeative, as the solvent flows through an immobile network of jammed droplets. The distinct behaviour of the droplet and solvent components in the suspension is instructive, and shows that the composite material behaves in a more complex way than what would be predicted for an ideal single-phase yield-stress fluid. In our conserved model, yielding can therefore be viewed as a continuous transition between a permeation regime with jammed amorphous droplets where solvent flows mainly around them, and a flowing ordered phase. In the latter phase, the flow is plug-like~\cite{SI,Foglino2017}, as found experimentally for colloidal suspensions in a pressure-driven flow~\cite{Isa2007}.

It is interesting to contrast the behaviour we have just discussed with that of the non-conserved model, where evaporation and condensation effects 
are included. Surprisingly, replacing strict area conservation with a soft constraint leads to a complete loss of yield-stress behaviour (Figs.~\ref{fig2}c,d). In the non-conserved model, droplets flow at any value of the forcing, however small, so that it is not possible to define a yield stress. While a yielding-like behaviour can still be observed as a smooth crossover, there is no longer a singularity in the droplet velocity curve (Fig.~\ref{fig2}c). %The throughput flow is this time linear (Fig.~\ref{fig2}(c)). 
An analysis of the area of each droplet show that the droplet motion is accompanied by area oscillations whose magnitude in controlled by $\lambda$, signalling that motion occurs via evaporation-condensation (inset of Fig.~\ref{fig2}c). The behaviour of the throughput flow mirrors that of the droplet velocity in this non-conserved model (Fig.~\ref{fig2}d) %\textcolor{red}{I would stress that evaporation-condensation is the leading mechanism at the base of the dynamics}.

%The yield-stress behaviour is apparent from the plot of the mean droplet velocity versus forcing $ \langle v_y \rangle_d = \frac{1}{N} \sum_i \int_\Omega d\bm{r} \vartheta_H(\phi_i-\frac{\phi_0}{2}) v_y $ in Fig.~\ref{fig2}b, being $\vartheta_H(\cdot)$ the Heavside theta function and $\Omega$ the system area. yielding can be seen as a continuous transition (or a super-critical Hopf bifurcation) in our model. Close to criticality, the mean droplet speed behaves as $\langle v\rangle \sim (\Delta p - \Delta p_c)^{\beta}$, with $\beta\simeq XXX$. The phenomenology is reminiscent of that of the Prandtl-Thomlinson model, for which $\beta=1/2$, which describes a particle in a dashboard potential, and provides a simple microscopic model for dry friction~\cite{Popov2017}.

%Interestingly, even in the phase in which droplets are at rest, the underlying solvent flows (Fig.~\ref{fig1}c and Fig.~S1) 
%Measuring the throughput flow \textcolor{red}{$Q=\int d\bm{x} v_y$} as a function of $\Delta p$, we find that this is non-zero at all $\Delta p$, and that there is a well-defined linear regime at small forcing, which corresponds to a high but finite effective viscosity -- the latter quantity can be defined by dividing the throughput flow of a Newtonian fluid with viscosity $\eta_0$ by the throughput flow of the fluid in the suspension. The flow at $0<\Delta p<\Delta p_c$ can be classified as permeative, as the solvent flows through an immobile network of jammed droplets (\textcolor{red}{The packing fraction is below $\phi_G$, can we speak of jamming?}). The distinct behaviour of the droplet and solvent components in the suspension is instructive, and shows that the composite material behaves in a more complex way than what would be predicted for an ideal single-phase yield-stress fluid. In our conserved model, yielding can therefore be viewed as a continuous transition between a permeation regime with jammed droplets where solvent flows around and through them, and a flowing phase.

%Fig. 1: yield-stress behaviour in conserved-model
%A: snapshot of system before and after yielding
%B: average velocity of droplets as a function of forcing - yield-stress behaviour
%C: throughput flow or viscosity as a function of forcing - no yield-stress behaviour?

%Supplementary Movies 1 (before) and 2 (after) yielding

More insight into the fundamental difference between the conserved and non-conserved models can be gained by analysing the behaviour of a single droplet at a solid wall under an external forcing, and with neutral wetting boundary conditions (Fig.~S5 and Suppl. Movie 3). While in the conserved model the droplet sticks to the wall and requires a finite forcing to start moving, in the non-conserved model evaporation and condensation provide another pathway for contact line motion~\cite{Bonn2009}, and the droplet drifts along the wall for any value of the forcing.  Therefore, besides the presence of a percolating network of droplet overlaps (Fig.~\ref{fig1}), the existence of a well-defined yielding transition also requires a suitable behaviour of droplets close to the wall. % for droplets at the wall, resulting in a finite force needed to depin them from there. %Failure to provide a stick boundary conditions, for instance due to evaporation-condensation phenomena, transforms the yielding transition into a crossover.\textcolor{red}{Qui bisogna essere più cauti  facciamo un doppio salto sulla questione del pinning depinning }
%The emerging picture is therefore the following. For yield-stress behaviour to occur in our suspensions we need two ingredients. First, the droplets need to stick to the wall without drifting for sufficiently small forcing: this requires neutral wetting and strict area conservation. Second, the overlaps (or contacts) between different droplets need to form a percolating network of droplets for the whole suspension to be jammed.

\begin{figure}[t!]
  \includegraphics[width=\columnwidth]{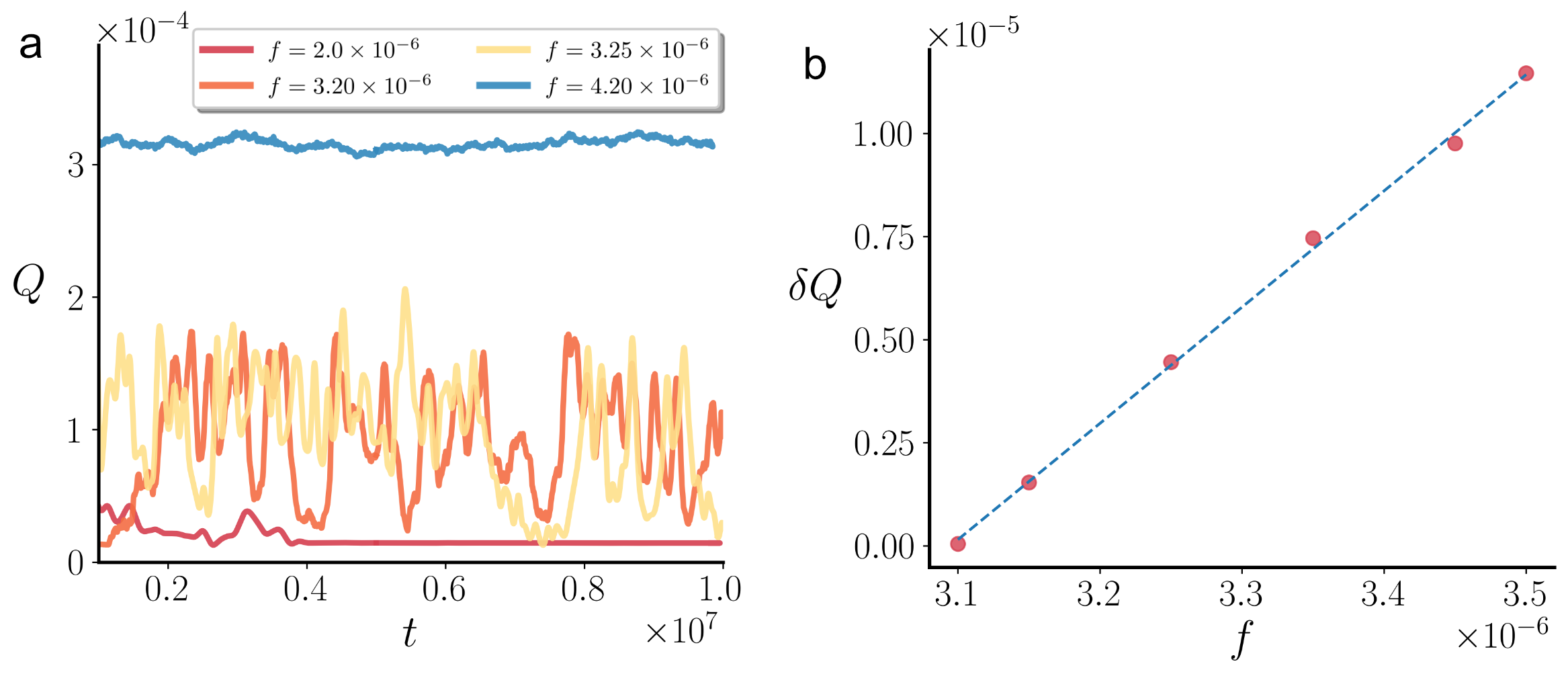}
  \caption{\textbf{Oscillations near the yielding transition.} \textbf{(a)} Throughput flow versus time, for the conserved model, for different values of $f$ near $f_c=3.15\times 10^{-6}$. \textbf{(b)} Plot of the variance of the oscillations as a function of $f$.} %\Livio{Larger labels in panel (a)}}
  \label{fig4}
\end{figure}

To understand the microscopic mechanism underlying yielding in the conserved model more deeply, we now consider their dynamics close to $f_c$. Just after yielding, we find a ``stick-slip'' behaviour where the emulsion alternates between plug-like motion, where droplets flow, with stationary spells, where they are almost jammed (Suppl.~Movie 4). The throughput solvent flow and the average droplet velocity both show irregular oscillations over time %, between a state which is almost jammed and flows very slowly, and another one which flows smoothly 
(red and orange curves in Fig.~\ref{fig4}a). The average variance (or amplitude) of the stochastic oscillations increases with the forcing, and approaches zero at $f_c$ (Fig.~\ref{fig4}b). This behaviour is reminiscent of that found in velocity oscillations of colloidal glasses close to the yielding transition~\cite{Isa2009}. There are indeed some key qualitative analogies between the two cases. In both systems, the non-flowing and flowing states subtly differ in the typical particle configuration. In our non-flowing emulsions, overlaps between droplets abound and create a nearly percolating chain through the system, just like frictional contacts for hard-sphere colloids. Instead, in the flowing states there are gaps between most particles~\cite{Isa2009}.
Analysing the dynamics in more details reveals an important distinction, though. In our system, instantaneous yielding events -- i.e., transitions from jammed to flowing states -- are typically accompanied by a sudden change in behaviour in the deformation free energy ($\mathcal{F}_{\textrm{def}}=\frac{K}{2}\int  \sum_{i}\left(\nabla\phi_i\right)^2 dy dz $, see~\cite{SI}). This suggest that yielding in our deformable suspensions requires a transient change in droplet shape,  which is instead essentially fixed for colloids. %so that having a finite $\gamma$ is important for the physics of yielding in our system, whereas $\gamma\to\infty$ for hard colloids.

%The oscillations we find closely resemble those found in colloidal glasses close to the yielding transition~\cite{Isa2009}, and we suggest that the underlying mechanism is similar.  In other words, contact rearrangements, which can be viewed as frictional interactions between droplets, are at the basis of the stochastic oscillations in Fig.~\ref{fig4}. %former state is reminiscent of frictional interactions between colloids, which are known to be crucial for their rheology -- for instance, to give discontinuous shear thickening~\cite{Wyart2014}.

\begin{figure}[t!]
  \includegraphics[width=1.0\columnwidth]{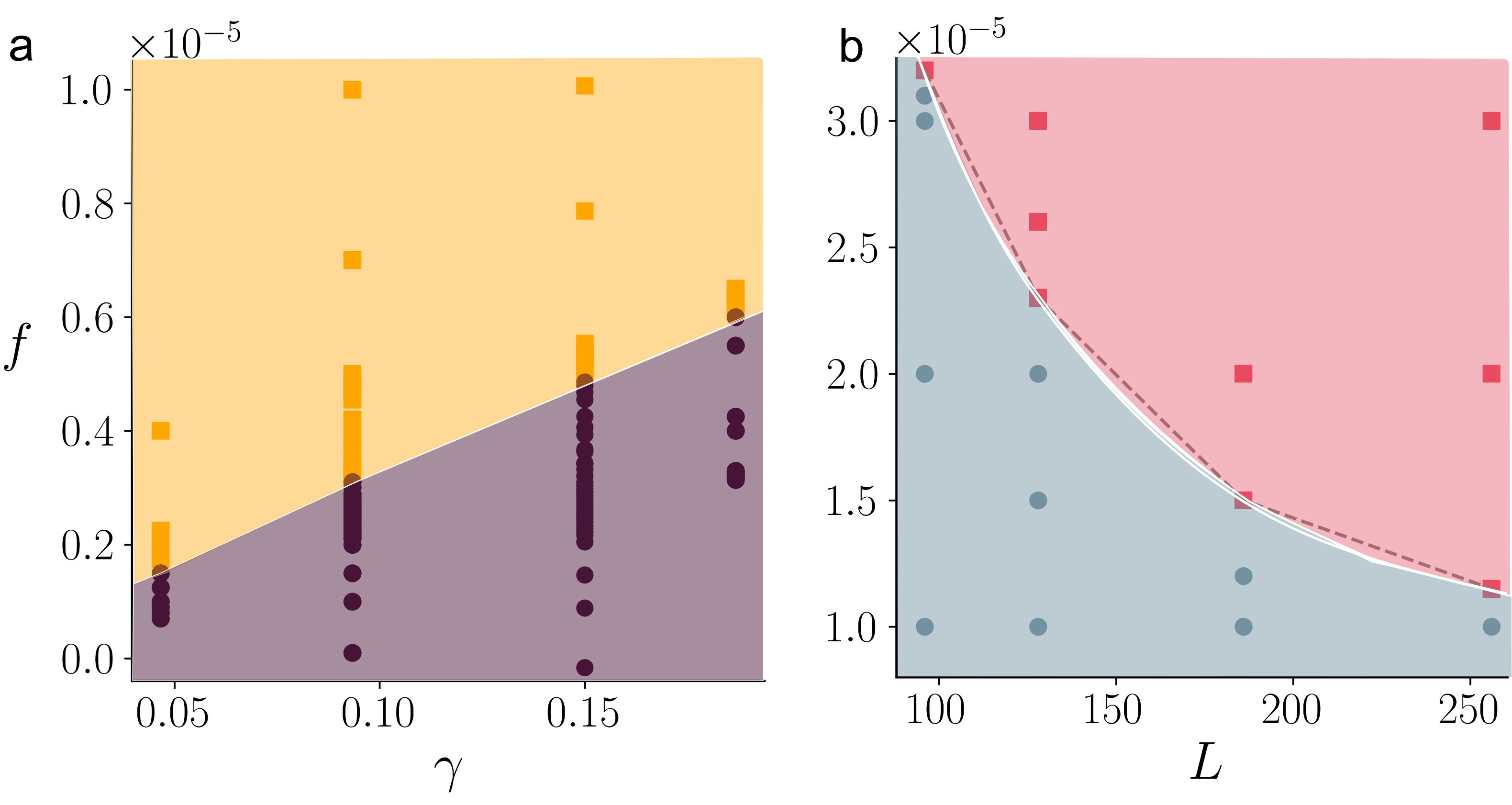}
  \caption{\textbf{Yielding phase diagram.} \textbf{(a)} Phase diagram as a function of body-force $f$ and surface tension $\gamma$.  Orange squares: flowing systems; purple circles: non-moving states.  \textbf{(b)} Phase diagram as a function of $f$  and system size $L$ plane.  Red squares: flowing systems; blue circles: non-moving states.  }\label{fig3}
\end{figure}

To verify that our qualitative mechanism for yielding through interfacial deformations is correct, we independently varied the parameters in Eq.~\ref{freeenergy} to see how they affect the value of the critical forcing. We found that $f_c$ scales linearly with surface tension, $\gamma$ (Fig.~\ref{fig3}a), and interfacial width, $\xi$ (see Fig. S6). The only other parameters appreciably affecting $f_c$ are the system size $L$ and the droplet radius $R$: increasing either of these lengthscales leads to a decrease in $f_c$ (Fig.~\ref{fig3}b). Our data therefore suggest that a key dimensionless parameter may be the capillary number $Ca=f LR^2/(\gamma \xi)$, which was also empirically found to determine the physics of discontinuous shear thinning~\cite{Foglino2017,Foglino2018,Fei2020}. This can be viewed as an inverse Bingham number $\sigma_v/\sigma_y$, with $\sigma_v\sim fL$ the viscous stress and $\sigma_y\sim \gamma\xi/R^2$ an effective yield stress. The form of this dimensionless control parameter suggests that in order for the suspension to yield the external forcing has to overcome free energy barriers associated with changes in particle shape, whose cost increases with $\gamma$ and $\xi$.

In summary, we studied the rheology of a soft droplet suspension under pressure-driven flow. %and asked whether it behaves as a yield-stress fluid. 
We found that the droplets only start moving when the forcing they are subjected to exceeds a critical threshold, as in an ideal yield-stress fluid. However, unlike one such material, even when droplets are jammed the solvent flows through them via permeation, as in sheared cholesteric~\cite{cholestericpermeation} and smectic liquid crystals~\cite{elasticpermeation}, leading to an effective viscosity which depends on the system geometry and ceases to be a bulk property of the material. Yielding is accompanied by a morphological transition. The jammed phase is amorphous and the network of droplet-droplet contacts, or overlaps, percolates in the direction perpendicular to the wall, conferring rigidity to the system. In the flowing phase, droplets order and contact percolation is lost. Within this picture, overlaps play a qualitatively similar role to frictional contacts in hard colloids~\cite{Wyart2014}. In our case, though, the transition between the jammed and flowing phase requires a transient change in droplet shape. More quantitatively, yielding occurs for a sufficiently large value of an inverse Bingham number, controlling the balance between viscous and interfacial stresses. The mechanism is therefore similar to that determining discontinuous shear thinning at larger forcing~\cite{Foglino2017,Foglino2018,Fei2020}: the main difference is that at the yielding transition interfacial deformations are spatially localised and transient in time, whereas at the discontinuous shear thinning transitions they affect large portions of the system and occur at all times.
%More in general, the yielding transition in the suspension can be viewed as a continuous transition between an amorphous jammed material permeation flow and an ordered suspension with plug-like flow. 
%Yielding requires a subtle rearrangement of droplet-droplet overlaps such that these no longer form a percolating network perpendicular to the wall directions. Within this picture, overlaps play a qualitatively similar role to frictional contacts in hard colloidal particles~\cite{Wyart2014}. 
%Close to the yielding transition, there are large amplitude oscillations between a nearly jammed and a flowing state, which are strikingly similar to those observed in colloidal glasses under flow.
%Yield-stress behaviour in our system requires droplets at the wall to be stationary for small forcing. %Second, to create a non-flowing jammed phase, overlaps or contacts between droplets need to self assemble to form a percolating network. 
%In line with this picture, 
Strikingly, we predict the yield-stress behaviour can be completely eliminated in our model by allowing droplet areas to fluctuate, for instance due to evaporation/condensation phenomena. We hope that our results will stimulate experiments to directly test our predictions, such as the importance of permeation and the scaling of $f_c$.
%and to stimulate future experiments in emulsions to test the key importance of wall interactions and droplet-droplet contacts to yield-stress behaviour. Our results also suggest that it may be fruitful to extend theories of discontinuous shear thickening to deformable emulsions with a yield-stress. 
To assess the universality of our results, one could investigate the yielding transition in other materials, such as biological tissues~\cite{Bi2016,Chiang2016,Loewe2020}, red blood cell suspensions~\cite{Lazaro2014}, and liquid crystalline emulsions~\cite{lcemulsions}. 

\section{Acknowledgments}
\begin{acknowledgments}
The work has been performed under the Project HPC-EUROPA3 (INFRAIA-2016-1-730897), with the support of the EC Research Innovation Action under the H2020 Programme. Part of this work was carried out on the Dutch national e-infrastructure with the support of SURF through the Grant 2021.028 for computational time (L.N.C and G.N.).  We acknowledge funding from MIUR Project No. PRIN 2020/PFCXPE.
\end{acknowledgments}

%\bibliographystyle{unsrt}
%\bibliography{biblio}

\end{document}